# Hubble Law or Hubble-Lemaître Law? The IAU Resolution

Helge Kragh*

**Abstract**: The Hubble law, widely considered the first observational basis for the expansion of the universe, may in the future be known as the Hubble-Lemaître law. This is what the General Assembly of the International Astronomical Union recommended at its recent meeting in Vienna. However, the resolution in favour of a renamed law is problematic in so far as concerns its arguments based on the history of cosmology in the relevant period from about 1927 to the early 1930s. A critical examination of the resolution reveals flaws of a non-trivial nature. The purpose of this note is to highlight these problems and to provide a better historically informed background for the voting among the union's members which in a few months' time will result in either a confirmation or a rejection of the decision made by the General Assembly.

On August 30, 2018, at the 30th Meeting of the International Astronomical Union (IAU), members of the General Assembly decided by simple majority to support Resolution B4 proposed by the IAU Executive Committee. The resolution recommends that instead of referring to Hubble's law, "from now on the expansion of the universe [should] be referred to as the 'Hubble-Lemaître law'."[1] Exceptionally and for the first time in the history of the century-old history of the IAU (which was founded in 1919), it was decided in advance that the vote of the General Assembly should be supplemented by an electronic vote among all the approximately 13,500 individual members of the union. The final decision regarding the renaming of Hubble's law thus depends on the electronic vote, the result of which will be revealed only later this year.

Of course, the astronomical community is not obliged to follow the IAU recommendation whatever its final outcome. According to the IAU, the names approved by the union "represent the consensus of professional astronomers around the world and national science academies, who as 'Individual Members' and 'National Members,' respectively, adhere to the guidelines of the International Astronomical Union."[2] Whether, in the present case, a recommended change of name is accepted or not can only be known *a posteriori* and depends entirely on the astronomers and the authors of textbooks in astronomy and cosmology.

---

* Niels Bohr Institute, University of Copenhagen, Denmark. E-mail: helge.kragh@nbi.ku.dk.
[1] https://www.iau.org/static/archives/announcements/pdf/ann18029e.pdf; http://astronomy2018.cosmoquest.org/newspaper/should-hubbles-law-be-renamed-your-vote-counts/

All the data, contexts and information relating to the Hubble-Lemaître proposal have been known for several decades, including that Georges Lemaître in 1927 predicted a cosmic velocity-distance law and estimated a numerical value for the recession or expansion constant. In a book of 1996 I mentioned, as others may have done earlier, that "The famous Hubble law is clearly in Lemaître's paper [and] it could as well have been named Lemaître's law."[3] It is not very clear why the Hubble-Lemaître proposal only appeared at the 2018 General Assembly and not at an earlier date. But perhaps, better late than never. Nor is it clear, given that Lemaître's contribution preceded Hubble's by two years, why the proposal is "Hubble-Lemaître" rather than "Lemaître-Hubble." No arguments are given by the Executive Committee.

Resolution B4 argues its case primarily from considerations based on the history of cosmology, stating among the aims of the resolution not only "to honour the intellectual integrity of Georges Lemaître" but also "to inform the future scientific discourses with historical facts." The resolution consequently lists a number of such relevant facts or what are claimed to be facts. It is on the basis of the historical material appended to the resolution that the General Assembly supported it, and it is on the same basis that the astronomers will cast their electronic votes. It appears to me that at least some of the appended historical considerations are of dubious validity and hence that they provide a questionable and even illegitimate background for the voting. If the IAU wants to set "the historical record straight" – and this is what the General Assembly says in its electronic newspaper – the result is remarkably poor.

In short, the astronomical community deserves a better and more correct version of the relevant works of Hubble and Lemaître in the late 1920s than the one presented in the IAU resolution. There exists a considerable scholarly literature on the subject,[4] but in the present communication I merely want to point to a few specific issues mentioned either explicitly or implicitly in the IAU resolution. I believe that these and other issues are relevant with regard to the voting.

---

[2] https://www.iau.org/public/themes/buying_star_names/

[3] H. Kragh, *Cosmology and Controversy: The Historical Development of Two Theories of the Universe* (Princeton: Princeton University Press, 1996), p. 30.

First, with regard to the name "Hubble law" it is not quite clear what the resolution refers to, as it seems to mix up the original Hubble law with later usage of the term. The resolution refers correctly to Hubble's "discovery of the apparent recession of the galaxies" but then switches to its current usage as an expression of the cosmic expansion. What became the Hubble law was established by Hubble in 1929 and, with more data and much more convincingly, restated in a 1931 paper by Hubble and his assistant Milton Humason. Although stated as a linear velocity-distance relation (not a law), both Hubble and Humason were careful to point out that it was really a redshift-distance relation and that it did not, in itself, justify the idea of an expanding universe. As Hubble knew, as early as 1929 Fritz Zwicky, in a work containing one of the earliest references to Hubble's paper, had derived what he called "Hubble's relation" on the assumption of a static universe. Until the mid-1940s no astronomer or physicist seems to have clearly identified Hubble as the discoverer of the cosmic expansion. Indeed, when Hubble went into his grave in 1953 he was happily unaware that he had discovered the expansion of the universe.

As to the name "Hubble law" we are told in Resolution B4 that it became universally adopted soon after the publication of the Hubble-Humason paper, but this was far from the case.[5] Apart from a single example from the 1930s, in a paper by Arthur G. Walker,[6] it seems that references to Hubble's law only began to appear in the scientific literature during the early 1950s. Among the first to use the term was George Gamow in his widely read *The Creation of the Universe*.[7] It took another decade until "Hubble's law" appeared regularly in the astronomical literature, then invariably as an expression of the expanding universe and as part of the myth that the cosmic expansion was discovered by Hubble.

---

[4] E.g. J. North, *The Measure of the Universe: A History of Modern Cosmology* (London: Oxford University Press, 1965); R. W. Smith, *The Expanding Universe: Astronomy's 'Great Debate' 1900-1931* (Cambridge: Cambridge University Press, 1982); Kragh, *Cosmology and Controversy* (ref. 3); J.-P. Luminet, *L'Invention du Big Bang* (Paris: Editions du Seuil, 2004); H. Nussbaumer and L. Bieri, *Discovering the Expanding Universe* (Cambridge: Cambridge University Press, 2009).
[5] H. Kragh and R. W. Smith, "Who discovered the expanding universe?" *History of Science* **41** (2003): 141-162.
[6] A. G. Walker, "Distance in an expanding universe," *Monthly Notices of the Royal Astronomical Society* **94** (1933): 159-167.
[7] G. Gamow, *The Creation of the Universe* (New York: Viking Press, 1952), p. 37; P. Couderc, *The Expansion of the Universe* (London: Faber & Faber, 1952), pp. 108-110.

It is worth mentioning that until about 1960 Humason's name often appeared together with Hubble's and that the term "Hubble-Humason law" can be found in several books and articles. Remarkably, Lemaître was one of the very first to refer to "Hubble's law," which he unambiguously credited to the American astronomer and not to himself. In a book review of 1950 he mentioned his calculation of the expansion constant dating from 1927 and then continued, "Naturally, before the discovery and study of galactic clusters, there could be no question of establishing Hubble's law, but only to determine the coefficient."[8] That is, he indirectly suggested that whereas he had more than a share in Hubble's constant, his role in the law was limited to a prediction without sufficient observational confirmation.

It has been known for some time that the English translation that Lemaître prepared of his ill-fated paper in *Annales de Société Scientifique de Bruxelles* left out important parts relating to his derivation of the expansion constant.[9] However, to state with the IAU that Lemaître omitted "the section in which he derived the expansion" or that he "edited out the expansion equations" is to go too far, I think. The 1931 version in *Monthly Notices* did contain the relevant equations including an approximate redshift-distance relation and also a value for the expansion constant (ca. 625 km/s/Mpc) based on what cryptically was said to be "a discussion of available data." Unfortunately the English translation offered no hints of what these data were and it also did not refer to the French original published four years earlier.

Was Hubble aware of Lemaître's theory of the expanding universe prior to his own work relating the redshifts of galaxies to their distances? He certainly knew the Belgian astrophysicist, who in 1925 listened to a talk Hubble gave in Washington D.C. and later the same year visited him at Mount Wilson to learn more about the redshift observations.[10] Now, here is what Resolution B4 writes about the next meeting between the two pioneers of cosmology: "Both Georges Lemaître and the American astronomer Edwin Hubble attended the 3rd IAU General Assembly in Leiden in July 1928 and exchanged views about the relevance of the redshifts vs distance observational data of the extragalactic nebulae to the emerging evolutionary model of the universe."

---

[8] G. Lemaître, *Annales d'Astrophysique* **13** (1950): 344-345, with partial translation in Kragh and Smith, "Who discovered the expanding universe?" (ref. 5), p. 147. See also D. Lambert, *The Atom of the Universe: The Life and Work of Georges Lemaître* (Cracow: Copernicus Center Press, 2015), where it is stated on p. 139 that "It would be more accurate to call it [Hubble's law] the 'Hubble-Lemaître law'."

[9] M. Livio, "Lost in translation: Mystery of the missing text solved," *Nature* **479** (2011): 171-173.

[10] Lambert, *The Atom of the Universe* (ref. 8), pp. 117-119.

To the best of my knowledge this is just wrong. Hubble was indeed a delegate at the Leiden General Assembly and a member of the IAU Commission on Nebulae and Star Clusters. Also Lemaître, who since 1925 had been a member of the IAU, was in Leiden, but was assisting in the organizational work and not participating as a delegate.[11] It is perhaps natural to assume that he would have contacted Hubble or vice versa, but there is no documentary evidence I know of that such a meeting or conversation actually took place. Noting that Lemaître's 1927 value of the recession constant was close to Hubble's value of 1929 (ca. 500 km/s/Mpc), the eminent cosmologist James Peebles suggested in 1971 that "there must have been communication of some sort between the two."[12] No such communication is recorded and the exchange of views cited by the IAU resolution seems to be a speculation unsupported by facts.

To make things worse, Resolution B4 refers in a note to two sources as evidence for the Hubble-Lemaître meeting in Leiden, but apparently without having checked the sources. An oral interview with Humason is cited "as reported by Sidney van den Bergh" in a paper of 2011. This paper, by a distinguished Canadian astronomer and former Vice-President of the IAU, does not mention either the interview with Humason or the Leiden conference, but is concerned only with the differences between Lemaître's French paper of 1927 and his English paper of 1931.[13] As to the Humason interview it was conducted by Bert Shapiro in about 1965 and later incorporated in the Oral History Interview series of the American Institute of Physics. All that Humason said about the Leiden meeting was that "The velocity-distance relationship started after one of the IAU meetings, I think it was held in Holland."[14] There is not a word in the interview about Lemaître. What is going on? It is hard to avoid the conclusion that on this point the IAU resolution comes close to fabricating evidence or is at least sloppy history. I am reminded of an early paper on Lemaître's contributions to cosmology, in which Peebles quipped, "Physical scientists have a healthy attitude toward the history of their subject: by and large we ignore it."[15]

---

[11] G. E. Christianson, *Edwin Hubble: Mariner of the Nebulae* (New York: Farrar, Straus and Giraux, 1995), p. 187; C. H. Gingrich, "The International Astronomical Union at Leiden," *Popular Astronomy* **36** (1928): 509-519; *Transactions of the International Astronomical Union, Vol. III. Third Assembly Held at Leiden 5-13 July 1918* (Cambridge: Cambridge University Press, 1929).

[12] P. J. E. Peebles, *Physical Cosmology* (Princeton: Princeton University Press, 1971), p. 8.

[13] S. van den Bergh, "The curious case of Lemaître's equation no. 24," *Journal of the Royal Astronomical Society of Canada* **105** (2011): 151.

The seminal contributions of Lemaître were for a long time neglected or not properly recognised by either the astronomical community or historians of science. Compared to the iconic Hubble, the Belgian priest and astrophysicist appeared as a somewhat shadowy figure. Incidentally, both scientists were nominated for the Nobel Prize in physics but in neither case successfully (Lemaître was also nominated for the chemistry prize, in 1955). Of some relevance for the IAU proposal, the nomination of Lemaître for the 1954 physics prize was for "his 1927 theoretical prediction of the expanding universe which was subsequently confirmed by the work of Hubble and Humason in the U.S.A."[16]

During the last few decades Lemaître's unique position in the history of modern cosmology has increasingly if slowly been recognised. As far as eponymies are concerned, Hubble has been richly rewarded, examples being not only the Hubble law and the Hubble constant but also and not least the Hubble Space Telescope. Lemaître's name too has entered scientists' vocabulary if less conspicuously and is largely unknown to the public. The basic equations for the dynamic universe are frequently called the Friedmann-Lemaître equations and the corresponding metric the Friedmann-Lemaître-Robertson-Walker (FLRW) metric; specialists in general relativity refer to the Lemaître-Tolman-Bondi (LTB) solution of Einstein's field equations for a spherically symmetric system of dust particles. His name has more recently been associated with an automated supply spacecraft for the International Space Station known as ATV-5 or the Georges Lemaître ATV.[17]

There are very good reasons to honour Lemaître and draw attention to his important work in cosmology, but perhaps the one suggested by the IAU General

---

[14] https://www.aip.org/history-programs/niels-bohr-library/oral-histories/4686

[15] P. J. E. Peebles, "Impact of Lemaître's ideas on modern cosmology," in *The Big Bang and Georges Lemaître*, ed. A. Berger (Dordrecht: Reidel, 1984), pp. 23-30.

[16] H. Kragh, "The Nobel Prize system and the astronomical sciences," *Journal for the History of Astronomy* **48** (2017): 257-280.

[17] https://en.wikipedia.org/wiki/Georges_Lema%C3%AEtre_ATV

Assembly is not the best one. Lemaître unquestionably was the first to propose and justify that the universe is in a state of expansion, which he did not only theoretically but also by making use of empirical data. On the other hand, it is more questionable if he should be credited for the empirical redshift-distance law, a law which he himself ascribed to Hubble.[18] In any case, the decision as to change the name of the law or not needs to be based on reliable historical sources.

Scientific names are generally important and sometimes carry with them epistemic connotations and mental images that other names do not convey. They are more than just convenient labels. But in science some names are more important than others.[19] Eponymies associating laws, theories or equations with great scientists of the past largely serve a social and ideological purpose rather than a scientific purpose. It does not make a great deal of difference whether astronomers speak of the HR law or the Hertzsprung-Russell law, or whether physicists associate the uncertainty principle with Heisenberg's name or not. "What's in a name?" Shakespeare asked in *Romeo and Juliet*. "That which we call a rose by any other name would smell as sweet."

When the General Assembly of the IAU in 2006 decided to categorise Pluto as a dwarf planet and not a planet, there were scientific reasons for the controversial change of name. When it comes to scientific eponymies such as the Hubble law the reasons are social and often rooted in national or institutional considerations. If they are to be justified at all it must be by sound historical documentation and not by speculations of what might have happened in the past.

---

[18] "Hubble must therefore be considered the discoverer of this empirical law. But the law of receding galaxies is not the same as the expanding universe." Kragh and Smith, "Who discovered the expanding universe?" (ref. 5), p. 153.

[19] A. L. Caso, "The production of new scientific terms," *American Speech* **55** (1980): 101-110.